\documentclass{article}

\usepackage[letterpaper]{geometry}

\usepackage{amsmath}
\usepackage{amssymb}
\usepackage{hyperref}
\usepackage{cleveref}
\usepackage{graphicx}
\usepackage{multicol}
\usepackage{multirow}

\usepackage[noend,noline,linesnumbered]{algorithm2e}

\newcommand{\pathcover}{\mathcal{P}}
\newcommand{\tmp}{\mathcal{T}}
\newcommand{\flowG}{\mathcal{G}}
\newcommand{\flowV}{\mathcal{V}}
\newcommand{\flowE}{\mathcal{E}}

\newcommand{\resi}[2]{\mathcal{R}(#1, #2)}

\crefname{algocf}{alg.}{algs.}
\Crefname{algocf}{Algorithm}{Algorithms}

\begin{document}

\title{\Large Minimum Path Cover: The Power of Parameterization\thanks{This work was partially funded by the US Fulbright program, the Fulbright Finland Foundation, the Helsinki Institute for Information Technology (HIIT), the US National Science Foundation (award DBI-1759522), the European Research Council (ERC) under the European Union's Horizon 2020 research and innovation programme (grant agreement No.~851093, SAFEBIO), and the Academy of Finland (grants No.~322595, 328877).}}

\author{Manuel Cáceres\thanks{Department of Computer Science, University of Helsinki, Finland, \texttt{manuel.caceresreyes@helsinki.fi}.}
\and Brendan Mumey\thanks{Gianforte School of Computer Science, Montana State University, USA, \texttt{brendan.mumey@montana.edu}.}
\and Santeri Toivonen\thanks{Department of Computer Science, University of Helsinki, Finland, \texttt{santeri.toivonen@helsinki.fi}.}
\and Alexandru~I.~Tomescu\thanks{Department of Computer Science, University of Helsinki, Finland, \texttt{alexandru.tomescu@helsinki.fi}.}}

\date{}

\maketitle

\begin{abstract} \small\baselineskip=9pt 
Computing a minimum path cover (MPC) of a directed acyclic graph (DAG) is a fundamental problem with a myriad of applications, including reachability.
Although it is known how to solve the problem by a simple reduction to minimum flow, recent theoretical advances exploit this idea to obtain algorithms parameterized by the number of paths of an MPC, known as the \emph{width}. These results obtain fast~[M\"akinen et~al., TALG] and even linear time~[C\'aceres et~al., SODA 2022] algorithms in the small-width regime.

In this paper, we present the first publicly available high-performance implementation of state-of-the-art MPC algorithms, including the parameterized approaches. Our experiments on random DAGs show that parameterized algorithms are orders-of-magnitude faster on dense graphs. Additionally, we present new pre-processing heuristics based on transitive edge sparsification. We show that our heuristics improve MPC-solvers by orders-of-magnitude.
\end{abstract}

\section{Introduction}

\subsection{Motivation} 
\begin{figure}[t]
    \centering
    \includegraphics[width=0.7\textwidth]{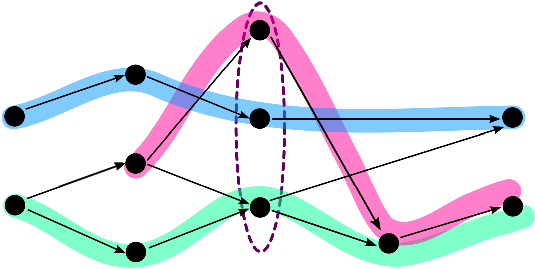}
    \caption{A DAG, an MPC shown as highlighted paths, and a maximum-sized set of non-reachable vertices in a dashed oval. The width of this graph is $k = 3$.}
    \label{fig:MPC}
\end{figure}

A minimum path cover $\pathcover$ (MPC) of a directed acyclic graph (DAG) $G = (V, E)$ is a minimum-sized set of paths covering $V$, that is, every vertex of $V$ is present in at least one path of $\pathcover$. Dilworth~\cite{dilworth1987decomposition} proved that the number of paths in such a set, namely the width $k$, equals the maximum number of pairwise non-reachable\footnote{A vertex $u$ reaches a vertex $v$ if there is a path from $u$ to $v$.} vertices. See \Cref{fig:MPC} for an illustration of these concepts. Later, Fulkerson~\cite{fulkerson1956note} showed that the problem of finding an MPC is polynomially solvable with a reduction to maximum matching in a bipartite graph encoding the reachability relation between the vertices.

Computing an MPC has many applications in many areas of computer science such as scheduling~\cite{colbourn1985minimizing,desrosiers1995time,bunte2009overview,zhan2016graph}, computational logic~\cite{bova2015model,gajarsky2015fo}, distributed computing~\cite{tomlinson1997monitoring,ikiz2006efficient}, evolutionary computation~\cite{jaskowski2011formal}, programming languages~\cite{kowaluk2008path},  databases~\cite{Jagadish90}, cryptography~\cite{mackinnon1985optimal}, and program testing~\cite{ntafos1979path}. In bioinformatics, MPCs are used on fundamental problems in pan-genomics~\cite{makinen2019sparse,cotumaccio2021indexing} and multi-assembly~\cite{eriksson2008viral,trapnell2010transcript,chang2015bridger}. Moreover, since an MPC covers the entire graph, it also encodes the reachability between the vertices, as formally shown by the constant-time reachability index of Jagadish~\cite{Jagadish90}. As such, MPCs are fundamental objects in the problem of reachability and the applications therein.

The results of Dilworth and Fulkerson were developed in the context of partially order sets (posets) where the input object corresponds to a transitive DAG. The problem was later defined on general DAGs (as presented in this manuscript) and solved by a simple and elegant reduction to minimum flow~\cite{ntafos1979path}, the \emph{folklore reduction}. In this reduction, a minimum flow is computed on a slightly modified graph $\flowG = (\flowV, \flowE)$, which is then decomposed to obtain the corresponding MPC. Hence, an MPC can be found in time $O(T_{MF}(\flowG) + ||\pathcover||)$, where $T_{MF}$ is the time to compute a maximum flow\footnote{In this reduction, the minimum flow problem can be easily reduced to maximum flow as we will explain later.} and $||\pathcover||$ is the total length of the paths in the computed MPC\footnote{The corresponding MPC can be decomposed from the flow in time $O(||\pathcover|| + |E|)$. We will explain this algorithm later.}. 

On the one hand, by using the recent breakthrough result on flows of Chen et~al.~\cite{chen2022maximum}, we can compute an MPC in almost-optimal $O(|E|^{1+o(|E|)} + ||\pathcover||)$-time. Although this is an impressive theoretical discovery, state-of-the-art flow algorithms rely on complex convex optimization techniques, and are far from being competitive in practice against current high-performance flow solvers (see e.g.~\cite{becker2016novel} for progress in this line of research). 

On the other hand, recent efforts further study the minimum flow reduction and develop algorithms parameterized by the width $k$, obtaining running times of $O(k(|V|+|E|)\log{|V|})$~\cite{felsner2003recognition,kowaluk2008path,makinen2019sparse} and the first parameterized linear time algorithm running in time $O(k^3|V|+|E|)$~\cite{caceres2022sparsifying} and later improved to $O(k^2|V|+|E|)$~\cite{caceres2023minimum}. Although these approaches are beaten in the large-width regime, they have practical potential as 1) they are simple combinatorial approaches, which also facilitates their implementation, 2) the expected width of random DAGs is known to be upper-bounded~\cite{barak1984maximal,simon1988improved} and 3) the width in several applications has been observed to be rather small~\cite{ma2023chaining,CJ23}.

\subsection{Contributions} 

In this work we present the first open source high-performance implementations of different MPC-solvers including:
\begin{itemize}
\setlength\itemsep{.05em}
    \item[--] The folklore reduction, which is compatible with all maximum flow and minimum cost flow solvers from the LEMON library~\cite{dezsHo2011lemon} as well as our own implementations of classical maximum flow algorithms.
    \item[--] The $O(k(|V|+|E|)\log{|V|})$-time algorithm~\cite{felsner2003recognition,kowaluk2008path,makinen2019sparse}, which is also compatible with all the flow solvers from the previous point.
    \item[--] The parameterized linear time algorithms $O(k^3|V|+|E|)$~\cite{caceres2022sparsifying}, $O(k^2|V|+|E|)$~\cite{caceres2022minimum}.
\end{itemize}

Our experiments on random DAGs show that 
the parameterized approaches are orders-of-magnitude faster than the folklore reduction on the fastest flow-solvers. In fact, our implementations of the parameterized approaches are able to compute MPCs on graphs with more than $10^8$ edges in less than $2$ minutes.
In particular, the parameterized linear time algorithms shine on dense and small-width instances and outperform all its competitors, running in less than $5$ seconds. 

We also present new fast pre-processing heuristics based on the concept of transitive sparsification~\cite{caceres2022sparsifying}. By removing transitive edges, our heuristics reduce the running time of solvers by up to an order-of-magnitude.\\

The rest of the paper is organized as follows. \Cref{sec:algorithms} explains the algorithms in our implementations, as well as some important implementation details. \Cref{sec:sparsification-heuristics} shows our proposed pre-processing heuristics based on transitive sparsification. \Cref{sec:experiments} presents our experimental setup and results.

\section{Flow-based MPC algorithms}\label{sec:algorithms}

All state-of-the-art MPC-algorithms are based on a simple and elegant reduction to minimum flow. Analogous to the maximum flow problem, in minimum flow~\cite{ciurea2004sequential} we are given a graph $\flowG = (\flowV, \flowE)$ with a source $s\in \flowV$ and a sink $t\in\flowV$, and \emph{demands} on the edges $d: \flowE \rightarrow \mathbb{N}_0$. The goal is to compute an $st$-flow (or just flow) $f^*: \flowE \rightarrow \mathbb{N}_0$ of minimum size $|f^*|$ (net flow exiting $s$), which satisfies \emph{flow conservation} (the flow entering and exiting a non-source nor sink vertex is the same) and respects the demands ($f^*(e) \ge d(e)$ for all edges). For a more formal definition of these concepts we refer to~\cite{ahujia1993network}.

\subsection{The Folklore reduction} 
\begin{figure}[t]
    \centering
    \includegraphics[width=0.7\textwidth]{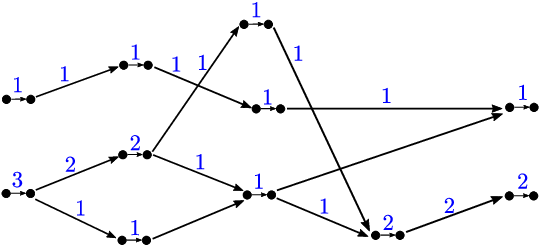}
    \caption{A feasible flow of size $4$ in the flow reduction of the graph shown in \Cref{fig:MPC}. Flow values are shown on top of the corresponding edges ($0$ if not present). Vertices $s$ and $t$ are not shown for simplicity. A decomposition of this flow produces a path cover with $4$ paths, i.e. not minimum.}
    \label{fig:minFlowReduction}
\end{figure}

The reduction from MPC to minimum flow has been discovered and re-discovered many times in the literature (see e.g.~\cite{caceres2022sparsifying}, here we use their notation), but it can be attributed to its first public appearance in the paper of Ntafos and Hakimi~\cite{ntafos1979path}. Given a DAG $G = (V, E)$, we build its \emph{flow reduction} as the pair $\flowG = (\flowV, \flowE)$, $d:\flowE \rightarrow \mathbb{N}_0$, where $\flowV$ contains two copies $v^{in}$, $v^{out}$ of each vertex $v \in V$ connected by an edge with demand $d(v^{in}, v^{out}) = 1$ (every other edge in $\flowE$ has demand $0$). The set $\flowV$ also contains a global source $s$ connected to every $v^{in}$ and a global sink $t$ connected from every $v^{out}$. Finally, $\flowE$ replicates $E$ by having an edge $(u^{out}, v^{in})$ for every edge $(u, v) \in E$. Note that $|\flowV| = O(|V|) , |\flowE| = O(|V|+|E|)$. A flow $f$ in this reduction corresponds to a path cover $\pathcover$ (not necessarily minimum) of $G$ with $|f|$ paths. See \Cref{fig:minFlowReduction} for an illustration of these concepts.

Each of these paths can be obtained by decomposing one unit of flow at a time from $f$. The decomposition can be naively performed with $f$ graph searches in total $O(|f|(|V|+|E|))$ time. However, we implemented a faster algorithm running in time $O(||\pathcover|| + |E|)$ described in \Cref{sec:decomposition}. 

As every path cover can be interpreted as a flow, an MPC corresponds to a minimum flow $f^*$ in this network. Minimum flow on the flow reduction can be reduced to maximum flow by also providing an initial flow $f$, that is, a path cover. For every edge $e \in \flowE$, we place it in the maximum flow instance only if $f(e) > d(e)$, in which case we define its capacity to be $c(e) = f(e) - d(e)$. Moreover, for every edge $(u, v) \in \flowE$, we place its reverse edge $(v, u)$ in the maximum flow instance with capacity $c(u,v) = |f|$. It can be shown~\cite{makinen2019sparse} that if $f'$ is a maximum flow of this instance, then $f^* = f'-f$ is a minimum flow of $\flowG, d$.\footnote{Flow in reverse edges is interpreted as negative flow in the opposite direction.}

\begin{figure}[t]
    \centering
    \includegraphics[width=0.7\textwidth]{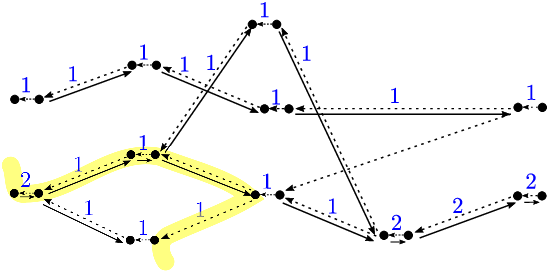}
    \caption{The residual graph of the flow shown in \Cref{fig:minFlowReduction}. Direct edges are shown as solid arrows, while reverse edges are shown as dashed arrows. A residual path is highlighted. The flow values obtained after using the residual path are shown on the corresponding edges.}
    \label{fig:residual}
\end{figure}

A more direct (yet equivalent) interpretation of the minimum flow problem given in~\cite{caceres2022sparsifying} defines the residual graph $\resi{\flowG}{f}$, by placing every \emph{reverse edge} (used to increase the flow in the opposite direction) and placing \emph{direct edges} whenever $f(e) > d(e)$ (used to decrease the flow). An $st$-path in $\resi{\flowG}{f}$ (\emph{residual path}) can then be used to decrease the flow size by one unit. See \Cref{fig:residual}.

In both interpretations of the problem, a minimum flow of the reduction can be obtained in time $O(|f|(|V|+|E|))$ by a simple Ford-Fulkerson approach~\cite{ford1956maximal}, which finds $O(|f|)$ residual paths. Since there is always a path cover that uses $|V|$ paths to cover every vertex (one path per vertex, we call this solution \emph{naive}), the previous approach runs in quadratic $O(|V|(|V|+|E|))$ time.

\subsection{Greedy solution}\label{sec:greedy}
Felsner et~al.~\cite{felsner2003recognition} proposed a greedy heuristic to compute a chain decomposition of a poset. They iteratively extract the longest chain of elements from the poset. They proved, with identical arguments to those of the greedy set cover logarithmic approximation~\cite{chvatal1979greedy}, that the number of chains extracted is bounded by $O(k\log{|V|})$. Later, Kowaluk et~al.~\cite{kowaluk2008path} showed that the same principle can be applied to general DAGs by finding the path covering the most uncovered vertices. They showed that these paths can be found in a DAG by a reduction to shortest path, which was later simplified by M\"akinen et~al.~\cite{makinen2019sparse} with a simple dynamic program. As such, computing the greedy solution and using it in the flow reduction produces an $O(k(|V|+|E|)\log{|V|})$-time algorithm for MPC.

\subsection{Parameterized linear time algorithms} Recently, C\'aceres et~al.~\cite{caceres2022sparsifying} proposed a slightly different method to compute an MPC using the flow reduction. In their method, the vertices are processed one by one in topological order~\cite{kahn1962topological,tarjan1976edge}\footnote{As topological sorting algorithms run in linear time, we assume that such order is given as input. In our implementation we use the faster DFS-based algorithm of Tarjan~\cite{tarjan1976edge}.} and an MPC of the (graph induced by the) already processed vertices is computed at each step. More specifically, if $\pathcover$ is an MPC of the first vertices in topological order and $v$ is the next vertex to process, the solution $\tmp = \pathcover \cup \{(v)\}$ is used as initial solution of the flow reduction. A nice property of $\tmp$ is that its size is either $|\pathcover|$ or $|\pathcover|+1$, and thus only one traversal of the residual suffices to obtain the MPC of the current iteration (or to check that $\tmp$ is an MPC).

By considering simple graph traversals of the residual, this approach runs in time $O(|V|(|V|+|E|))$. However, C\'aceres et~al. combine \emph{transitive sparsification} of edges with a special \emph{layered traversal} of the residual to obtain a linear dependency in the number of edges and a factor $k^3$ dependency in the number of vertices for a total running time of $O(k^3|V|+|E|)$.\\

\begin{algorithm}[t]
\DontPrintSemicolon
\SetKwFunction{FMain}{layeredTraversal}
  \SetKwProg{Fn}{Function}{:}{}
  \Fn{\FMain{$\flowG, f, \ell, v$}}{
    \vspace{.5em}
    $S \gets \{v^{in}\}$ \tcp{Visited vertices}
    \tcp{For each $j\in\{0,\ldots, |f|\}$}
    $Q_{j} \gets \{u^{out} \mid (u, v) \in E \land \ell(u^{out}) = j\}$\;
    \For{$j \gets |f|$ down to $0$}{
        \While{$Q_j \neq \emptyset$}{
            Remove $u$ from the front of $Q_j$\;
            $S \gets S \cup \{u\}$\;
            \For{$(u, v) \in N^+(u)$ in $\resi{\flowG}{f}$}{
                \If{$v = t$}{
                    \Return Residual path $D, S$
                }
                \If{$v\not\in S$}{
                    Add $v$ to the back of $Q_{\ell(v)}$
                }
            }
            
        }
    }
    \Return $\emptyset, S$ 
 }
 \caption{Layered traversal for the search of a residual path in $\resi{\flowG}{f}$. The algorithm returns a residual path (if one is found) as well as the set of visited vertices during the traversal.\label{alg:layeredTraversal}}
\end{algorithm}

\noindent \textbf{Layered traversal.} The algorithm assigns a level $\ell: \flowV \rightarrow \{0, \ldots, |\pathcover|\}$ to every vertex in the flow reduction. The level assignment maintains the property that paths in the residual graph are sequences of vertices with non-increasing levels, which allows to perform the traversal for the search of a residual path in a layered\footnote{We call \emph{layer} to the vertices in the same level.} manner. The residual graph is traversed from the highest reachable layer until the lowest reachable layer (or until a residual path is found). To perform this layered traversal, the algorithm uses $|\pathcover|+1$ FIFO queues (one per layer), each of which performs a BFS from the highest-reachable layer down to the lowest-reachable layer. \Cref{alg:layeredTraversal} shows the corresponding pseudocode of the traversal.

After the layered traversal finishes, the algorithm updates the flow and level assignment to maintain the algorithm's invariants. The flow is only updated if the traversal finds a residual path, in such a case, the flow in direct edges of the path is decreased by one and the flow in (the reverse of) reverse edges is increased by one, which decreases the total flow by one. As for the level assignment, if the lowest visited level is $l$, then all visited vertices change their their level to $l$ while the level of $v^{out}$ (second copy of the current vertex) is set to $l+1$. Additionally, if there is no flow from layer $l$ exiting directly to $t$, then the algorithm performs a \emph{merge} of layer $l$. By ease of explanation, we skip the explanation of the merge procedure and instead refer to the original publication~\cite{caceres2022sparsifying} and our code.\\

\begin{figure}[t]
    \centering
    \includegraphics[width=0.65\textwidth]{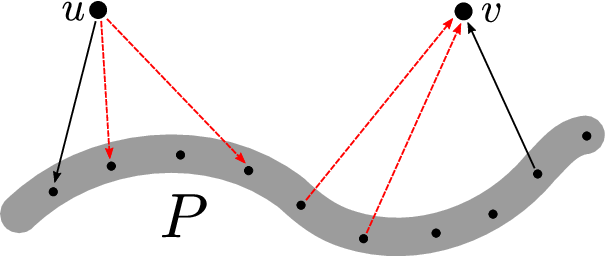}
    \caption{A path $P$ is highlighted (path edges were removed by simplicity). The figure shows (dashed arrows) the transitive edges incoming to a vertex $v$ as well as outgoing from a vertex $u$.}
    \label{fig:transitiveEdges}
\end{figure}  

\noindent \textbf{Transitive sparsification.} An edge $(u, v)$ is transitive if there is another path (different from the edge) from $u$ to $v$. Transitive edges can be removed from the DAG when computing an MPC, as removing these edges preserves the reachability relation between the vertices and hence the width. A \emph{transitive sparsification} is both a spanning subgraph with the same reachability relation as $G$ (some transitive edges might not be present) as well as the process to obtain such a subgraph. The $O(k^3|V|+|E|)$-time algorithm sparsifies the number of incoming edges to each vertex to $O(k)$. The authors use a simple idea first proposed by Jagadish~\cite{Jagadish90} on posets: if several incoming edges to $v$ come from the same path $P$, then all these edges, except maybe the last, are transitive. Conversely, if several outgoing edges from $u$ go to the same path $P$, then all these edges, except maybe the first, are transitive. See \Cref{fig:transitiveEdges}. The algorithm of C\'aceres et~al. uses this idea and the MPC from the previous iteration $\pathcover$ to sparsify the edges to the current vertex $v$ to at most $|\pathcover| \le k$. To perform this sparsification efficiently, the algorithm requires that every vertex stores the id of one path that contains such vertex, which is achieved by maintaining a path decomposition (an MPC $\pathcover$) of the minimum flow $f^*$.\\

\noindent \textbf{The $O(k^2|V|+ |E|)$-time algorithm.} 
The same authors later improved the running time of their algorithm~\cite{caceres2022minimum} by shaving a $k$-factor from the dependency on the number of vertices. They noted that is not necessary to maintain the ids of all paths containing a vertex (the MPC) during the algorithm, but that it suffices to maintain only one of those ids, and the MPC can be retrieved at the end by performing only one decomposition. To achieve this, they identified a set of \emph{antichain vertices} separating consecutive layers: as these vertices form an antichain, each of those must be covered by a different path, and in fact the algorithm covers each of these vertices with exactly one path (we refer to the original publication~\cite{caceres2022minimum} for details). As such, it suffices that every vertex points back to (one of) the corresponding antichain vertex on its layer, these pointers are called \emph{back links} ($bl$ in the pseudocode for short). As opposed to an entire decomposition, back links can be maintained by only decomposing the vertices in layer $l$ (lowest visited level), and then fixing the back links of vertices in lower layers (in constant time per vertex of level $> l$). \Cref{alg:k2} shows the pseudocode of one iteration of this algorithm.

\begin{algorithm*}[t]
\DontPrintSemicolon
\begin{multicols}{2}
        \tcp{Add $v$ to flow reduction}
        $\flowV \gets \flowV \cup \{v^{in}, v^{out}\}$\;  
        $\flowE \gets \flowE \cup \{(s, v^{in}), (v^{in}, v^{out}), (v^{out}, t)\}$\;
        \vspace{.5em}
        \tcp{Create initial solution $\tmp$}
        $f^*(s, v^{in}), f^*(v^{in}, v^{out}), f^*(v^{out}, t) \gets 1$\;
        \vspace{.5em}
        \tcp{Sparsify edges incoming to $v$}
        $\texttt{sur} \gets (\bot)^{|f^*|}$\;
        \For{$u\in N^-(v)$ in $G$}{
            $\texttt{sur}[id(bl(u))] \gets \max(\texttt{sur}[id(bl(u))], u)$
        }
        \For{$u \in \texttt{sur}, u\neq \bot$}{
            \tcp{Add $(u,v)$ to flow reduction}
            $\flowE \gets \flowE \cup \{(u^{out}, v^{in})\}, f^*(u^{out}, v^{in}) \gets 0$\;
        }
        \vspace{.5em}
        \tcp{Layered traversal with \Cref{alg:layeredTraversal}}
        $D, S \gets layeredTraversal(\flowG, f^*, \ell, v)$\;
        $l \gets \min\{\ell(u) \mid u \in S\}$\;
        \If(\tcp*[h]{Update flow values}){$D \neq \emptyset$}{
            \For{$(u,w) \in D$}{
                \lIf{$(u,w) \in \flowE$}{
                    $f^*(u,w) \gets f^*(u,w) - 1$
                }\lElse{
                    $f^*(w,u) \gets f^*(w,u) + 1$
                }
            }
        }\Else(\tcp*[h]{Update path and backlink info}){
            $id(v) \gets |f^*|, bl(v) \gets v$\;
        }
        \vspace{.5em}
        \tcp{Update levels}
        $\ell(v^{out}) \gets l+1$\;
        $\ell(u) \gets l$, for $u \in S$\;
        \vspace{.5em}
        \tcp{Decompose vertices in layer $l$}
        \For{\emph{Decomposed path }${}_uP_{w}$}{
            \tcp{Update links and path ids}
            $id(w) \gets id(u)$\;
            \For{$x \in {}_uP_{w}$}{
            \lIf{$x \neq u$}{
                    $bl(x) \gets u$\
                }
                $nl(x) \gets w$
            }
        }
        \vspace{.5em}
        \tcp{Fix backlinks in higher layers}
        \For{$u \in V, \ell(u^{in}) > l \lor \ell(u^{out}) > l$ in top. order}{
            \lIf{$\ell(bl(u)^{out}) \neq \ell(u^{in})$}{
                $bl(u) \gets nl(bl(u))$
            }
            \lIf{$\ell(u^{in}) \neq \ell(u^{out})$}{
                $id(u) \gets id(bl(u))$
            }
        }
        \vspace{.5em}
        \If{No flow from layer $l$ to $t$}{
            \tcp{Merge of layer $l$}
        }
\end{multicols}
\vspace{1em}
\caption{One iteration of the $O(k^2|V|+|E|)$-time algorithm. The pseudocode shows the steps of the algorithm when processing vertex $v$. Function $id$ returns the path id stored in an antichain vertex. Function $bl$ returns the corresponding antichain vertex (back link). Notation ${}_uP_{w}$ indicates the decomposed path connecting antichain vertices $u$ and $w$. For more details we refer to the original publication~\cite{caceres2022minimum} and our code.\label{alg:k2}}
\end{algorithm*}

\subsection{The decomposition algorithm}\label{sec:decomposition} As mentioned earlier, the last step of all MPC-solvers, as well as intermediate steps of the parameterized linear time algorithms, require to decompose the flow $f^*$ into an MPC. A naive solution extracts one path at a time in total $O(|f^*|(|V|+|E|))$ running time. We instead implement an algorithm that runs in time $O(||\pathcover|| + |E|)$\footnote{Note that $||\pathcover|| = O(|f^*|\cdot|V|)$, and thus this approach removes a factor $|f^*|$ from $|E|$, which is significant on non-sparse graphs.}. Such an improvement was first described by Kogan and Parter~\cite{kogan2022beating}. Here we use the version of C\'aceres~\cite{caceres2023minimum}. The algorithm first removes all $0$-flow edges in time $O(|E|)$ and then processes the vertices in topological order. When processing vertex $v$, it iterates through each in-neighbor $u$ and places $v$ after $u$ in $f^*(u,v)$ different paths. As such, the total running time equals $\sum_{v \in V, (u,v) \in E} f^*(u,v) = O(||\pathcover||)$.

\section{Pre-processing sparsification heuristics}\label{sec:sparsification-heuristics}

In this section, we present two transitive sparsification heuristics. Recall that a transitive sparsification removes transitive edges, making the input graph sparser. These heuristics are intended to be used as pre-processing steps of MPC-solvers to speed up their computation. As such, we ensure that their running time is upper-bounded by the running time of state-of-the-art solvers. Both of our heuristics use paths to sparsify the incoming/outgoing edges to/from a vertex as done by the parameterized linear time algorithms.\\

\begin{algorithm}[t]
\DontPrintSemicolon
\SetKwFunction{FMain}{dfsSp}
\SetKwProg{Fn}{Function}{:}{}
    \tcp{Global variables}
    \vspace{.5em}
    $E' \gets \emptyset$ \tcp{Transitive sparsification}
    $S \gets \emptyset$ \tcp{Visited vertices}
    $\texttt{dfs\_pre}[v] \gets 0$ for $v \in V$\;
    $\texttt{next\_pre} \gets 1$ \;
    $\texttt{last\_reach}[v] \gets 0$ for $v \in V$\;
    \vspace{.5em}
    \For{$v \in V$ in top. order}{
    \lIf{$v \not\in S$}{
    \FMain{v}
    }
  }
  \Return $G' = (V, E')$\;
  \vspace{1em}
  \Fn{\FMain{$v$}}{
        $S \gets S \cup \{v\}$\;
        $\texttt{dfs\_pre}[v] \gets \texttt{next\_pre} $\;
        $\texttt{next\_pre} \gets \texttt{next\_pre} + 1 $\;
        \vspace{.5em}
        \For{$w \in N^+(v)$ in top. order}{
            \If{$w\not\in S$}{
                \FMain{w}\;
            }
            \If{\emph{$\texttt{last\_reach}[w] < \texttt{dfs\_pre}[v]$}}{
                $E' \gets E' \cup \{(v,w)\}$\;
                $\texttt{last\_reach}[w] \gets \texttt{dfs\_pre}[v]$\;
            }
        }
  }
  \vspace{1em}
 \caption{DFS sparsification heuristic.}\label{alg:dfsSparsification}
\end{algorithm}

\noindent \textbf{DFS sparsification.} Our first sparsification heuristic uses the root-to-leaf paths of a DFS-spanning tree. A first naive implementation of this idea processes each of these paths to sparsify the incoming edges. However, this approach runs in time proportional to the total length of the root-to-leaf paths, which can be $\Omega(|V|^2)$. Instead, our algorithm runs in time $O(|V|+|E|)$ as it is implemented on top of a normal recursive DFS traversal. \Cref{alg:dfsSparsification} shows the corresponding pseudocode.

The main idea behind this algorithm is to use the DFS recursion itself as DFS paths. For this, it stores the preorder of each vertex visited (in \texttt{dfs\_pre}) as well as the maximum preorder value observed of an in-neighbor (in \texttt{last\_reach}). When processing an edge $(v, w)$ (after traversing $w$), if the observed preorder value of an in-neighbor of $w$ is bigger than the preorder of $v$ ($\texttt{last\_reach}[w] > \texttt{dfs\_pre}[v]$), then the edge is transitive and it is not added to the sparsification, as there is a vertex further down the DFS-tree also with an edge to $w$ (the one with preorder value $\texttt{last\_reach}[w]$). Conversely, among all vertices in a DFS root-to-leaf path with an edge to $w$, the only edge that is not sparsified is the one with the largest preorder value, that is, the one closer to the leaf.\\

\begin{algorithm}[t]
\DontPrintSemicolon
        $E' \gets E$ \tcp{Transitive sparsification}
        $\pathcover \gets \emptyset$ \tcp{Greedy path cover}
        \vspace{.5em}
        \While{$\pathcover$\emph{ is not a path cover}}{
            $P^* \gets$ path of $(V, E')$ with most uncovered\;
            $\pathcover \gets \pathcover \cup \{P^*\}$\;
            $R \gets \emptyset$ \tcp{In-neighbors to $P^*$}
            \For{$v \in P^*$}{
                \For{$u\in N^-(v)$ in $(V, E')$}{
                    \If{$u \in R \land (u,v) \not\in \pathcover$}{
                        $E' \gets E' \setminus \{(u,v)\}$\;
                    }
                    $R \gets R \cup \{u\}$\;
                }    
            }
        }
  \Return $G' = (V, E'), \pathcover$\;
  \vspace{1em}
 \caption{Greedy sparsification heuristic.}\label{alg:greedySparsification}
\end{algorithm}

\noindent \textbf{Greedy sparsification.} Our second sparsification heuristic also outputs the greedy initial solution explained in \Cref{sec:greedy}. It uses the $O(k\log{|V|})$ paths from the greedy solution to sparsify outgoing edges. As such, this heuristic sparsifies the edges to $|E'| = O(k|V|\log{|V|})$. Since this algorithm computes the greedy solution, its worst-case running time is also $O(k(|V|+|E|))$. However, we implemented a practical improvement where each extracted path is immediately used to sparsify, and thus the following paths are extracted from a sparser graph. \Cref{alg:greedySparsification} shows the corresponding pseudocode. Note that the algorithm does not sparsify an edge if this is present in the greedy path cover, however, there are at most $||\pathcover|| = O(k|V|\log{|V|})$ such edges.

\section{Experiments and Results}\label{sec:experiments}

\subsection{Implementations}
We implemented different flow-based MPC algorithms. The code was written in C++ and it can be found at  \url{https://github.com/algbio/PerformanceMPC} under the GNU General Public License v3.0. Our code is compatible with all maximum flow and minimum cost flow solvers from the LEMON library~\cite{dezsHo2011lemon}, which are known to be the fastest publicly available flow solvers~\cite{kovacs2015minimum}. However, for cleaner and fairer comparison of the approaches, in this work we use own implementations of the following well-known maximum flow solvers:
\begin{itemize}
\setlength\itemsep{.05em}
    \item \texttt{DFS}: Implements a Ford-Fulkerson~\cite{ford1956maximal} approach that finds residual paths using depth-first search.
    \item \texttt{BFS}: Implements Edmonds-Karp algorithm~\cite{edmonds1972theoretical}, which finds residual paths using breath-first search.
    \item \texttt{Blocking}: Implements Dinitiz' algorithm~\cite{dinitz2006dinitz}, which uses \emph{blocking flows}.
\end{itemize}

All our maximum flow-based solvers can start from one of the following initial solutions (path covers):
\begin{itemize}
\setlength\itemsep{.05em}
    \item \texttt{naive}: $|V|$ paths, each covering exactly one vertex.
    \item \texttt{greedy}: $O(\log{|V|})$-approximation ($O(k\log{|V|})$ paths) based on greedy set cover~\cite{felsner2003recognition,kowaluk2008path,makinen2019sparse}.
\end{itemize}

After running the flow solver all our implementations run the same $O(||\pathcover|| + |E|)$-time decomposition routine to obtain the corresponding MPC $\pathcover$ (see \Cref{sec:decomposition}).\\

Our code also implements the parameterized linear time algorithms:

\begin{itemize}
\setlength\itemsep{.05em}
    \item \texttt{k3}: Implements the first parameterized linear time algorithm running in time $O(k^3|V|+|E|)$~\cite{caceres2022sparsifying}.
    \item \texttt{k2}: Implements a later improvement over \texttt{k3} running in time $O(k^2|V|+|E|)$~\cite{caceres2022minimum}. 
\end{itemize}

To the best of our knowledge there are no other publicly available fast MPC-solver's implementations. Most publicly available MPC-solvers use the slower reduction to bipartite maximum matching, and thus also need to compute the transitive closure\footnote{The densest spanning supergraph having the original graph as a transitive sparsification.}. M\"akinen et~al.~\cite{makinen2019sparse} were the first to implement the greedy-based approach, which was later improved by Ma et~al.~\cite{ma2023chaining} using Dinitz' algorithm for finding residual paths: these implementations correspond to our \texttt{DFS greedy} and \texttt{Blocking greedy}, respectively.\\

Finally, for all our MPC-solvers we also implemented (as optional pre-processing) our two sparsification heuristics from \Cref{sec:sparsification-heuristics}: \texttt{dfs-sp} and \texttt{greedy-sp}.

\subsection{Setup}
The experiments ran on an isolated Intel(R) Xeon(R) CPU E5-2670 @ 2.6 GHz with 64GB of RAM, running Almalinux 8.4 (64bit, kernel 4.18.0). The code was compiled using \texttt{gcc} version 8.5.0 with optimization flag \texttt{-O3}. We measure \texttt{user} time using the \texttt{sys/resource.h} Unix library. We report the average value of $10$ repetitions of each experiment. We used a timeout of $10$ minutes for each experiment.

\begin{table}[t]
\centering
\begin{tabular}{|c|c|c|c|c|c|}
\hline 
$M = |E|$ & Width $k$ \\ \hline
$2^{15}$ & 31,282 \\
$2^{16}$ & 22,586 \\
$2^{17}$ & 13,913 \\
$2^{18}$ & 7,418 \\
$2^{19}$ & 3,768 \\
$2^{20}$ & 1,922 \\
$2^{21}$ & 980 \\
$2^{22}$ & 494 \\
$2^{23}$ & 260 \\
$2^{24}$ & 134 \\
$2^{25}$ & 75 \\
$2^{26}$ & 39 \\
$2^{27}$ & 22 \\
\hline
\end{tabular}
\caption{Width $k$ for dataset \texttt{Random DAG} ($N = 50,000$) and different values of parameter $M$.}
\label{tab:datasets_width}
\end{table}

\begin{table}[t]
\centering
\begin{tabular}{|c|c|c|c|c|c|}
\hline 
\multirow{3}{*}{$M$} & \multicolumn{2}{|c|}{\texttt{Path Partition}} & \multicolumn{2}{|c|}{\texttt{Transitive Closure}} \\
& \multicolumn{2}{|c|}{$N = 50,000, K = 173$} & \multicolumn{2}{|c|}{$N = 20,000$} \\ \cline{2-5}
 & $|E|$ & $k$ & $|E|$ & $k$   \\ \hline
$2^{14}$ & --- & --- & 30,837 & 11,441\\
$2^{15}$ & 82,595 & 173 & 135,638 & 7,926\\
$2^{16}$ & 115,363 & 173 & 1,964,397 & 4,631\\
$2^{17}$ & 180,899 & 173 & 37,401,045 & 2,339\\
$2^{18}$ & 311,971 & 173 & 106,958,060 & 1,233\\
$2^{19}$ & 574,115 & 173 & 153,889,748 & 625\\
$2^{20}$ & 1,098,403 & 171 & 178,537,108 & 318\\
$2^{21}$ & 2,146,979 & 165 & 190,392,813 & 164\\
$2^{22}$ & 4,244,131 & 140 & 195,826,240 & 90\\
$2^{23}$ & 8,438,435 & 111 & 198,244,885 & 48\\
$2^{24}$ & 16,827,043 & 79 & 199,289,591 & 25\\
$2^{25}$ & 33,604,259 & 54 & 199,733,890 & 15\\
$2^{26}$ & 67,158,691 & 34 & --- & ---\\
$2^{27}$ & 134,267,555 & 21 & --- & ---\\
\hline
\end{tabular}
\caption{Number of edges $|E|$ and width $k$ for datasets \texttt{Path Partition} and \texttt{Transitive Closure} and different values of parameter $M$.}
\label{tab:datasets_edges_width}
\end{table}

\subsection{Datasets}
We use the following classes of DAGs.\\

\noindent\textbf{Random DAG.}
For a fixed value of $N$ and $M$, we generate a random DAG with $N$ vertices and $M$ edges. The generation procedure first fixes a topological order of the $N$ vertices. Then, it generates $M$ different pairs of vertices and interprets them as edges directed according to the topological order. We fix $N = 50,000$ and vary $M \in \{32,768 = 2^{15}, 2^{16}, \ldots, 2^{27} = 134,217,128\}$ to observe the behavior at different densities. This dataset corresponds to the random DAG model proposed by Barak and Erd\"os~\cite{barak1984maximal}. We use this dataset to compare general performance. 
\Cref{tab:datasets_width} shows the width of DAGs in this dataset. Note that the width decreases with the number of edges of the \texttt{Random DAG}. Indeed, the expected width of a \texttt{Random DAG} of parameters $N$ and $M$, is upper bounded by $O(\frac{\log{(p\cdot n)}}{p})$~\cite{simon1988improved}, where $p = \frac{M}{\binom{N}{2}}$.\\

\noindent\textbf{Path Partition.}
For a fixed value of $N$, $M$ and $K$, we generate the previously described \texttt{Random DAG} with $N$ vertices and $M$ edges. Then, we divide the $N$ vertices into $K$ parts by placing each vertex on a uniformly random chosen part. Finally, we add the corresponding $N-K$ edges (in topological order) so that each part is a path in the DAG. As such, the graph's width is at most $K$. We fix $N=50,000$, $K=173$ and vary $M \in \{32,768 = 2^{15}, 2^{16}, \ldots, 2^{27} = 134,217,128\}$ as before. We use this dataset to study the performance on small-width instances. \Cref{tab:datasets_edges_width} shows the number of edges and width of DAGs in this dataset.\\

\noindent\textbf{Transitive Closure.}
For a fixed value of $N$ and $M$, we generate a \texttt{Random DAG} with $N$ vertices and $M$ edges. Then, we compute its transitive closure. We fix $N = 20,000$ and vary $M \in \{16,384=2^{14}, \ldots, 2^{25} = 33,554,432\}$. We use this dataset to study the performance on posets and the behavior of transitive sparsification heuristics. \Cref{tab:datasets_edges_width} shows the number of edges and width of DAGs in this dataset. Note that the width distribution of \texttt{Random DAG} is not affected as adding transitive edges does not change the width.

\begin{figure*}[t]
    \includegraphics[width=\textwidth]{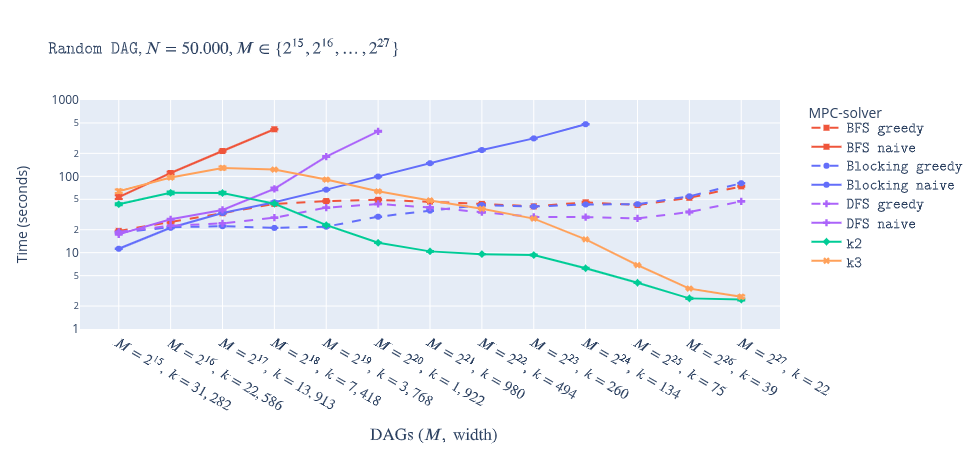}
    \caption{Running time of MPC-solvers in dataset \texttt{Random DAG}. Note the log-scale in the y-axis.}\label{fig:random_dag}
\end{figure*}

\begin{figure*}[t]
    \includegraphics[width=\textwidth]{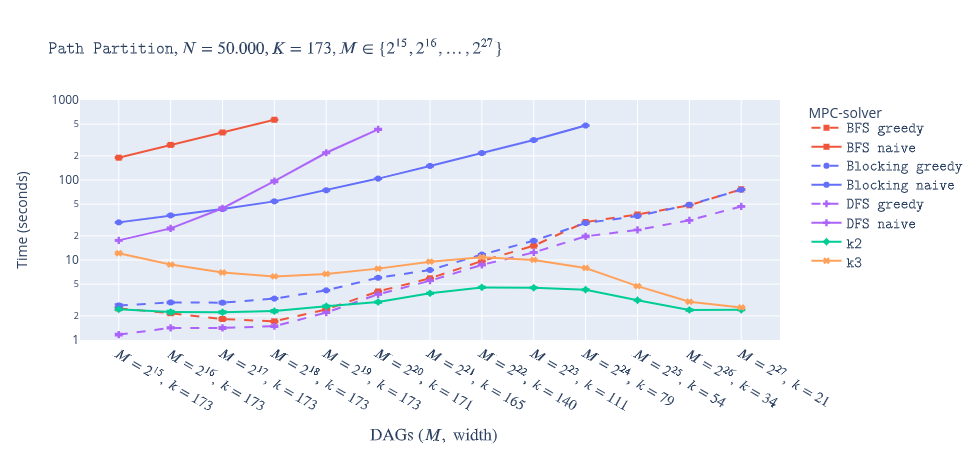}
    \caption{Running time of MPC-solvers in dataset \texttt{Path Partition}. Note the log-scale in the y-axis.}\label{fig:path_partition}
\end{figure*}

\begin{figure*}[t]
    \includegraphics[width=\textwidth]{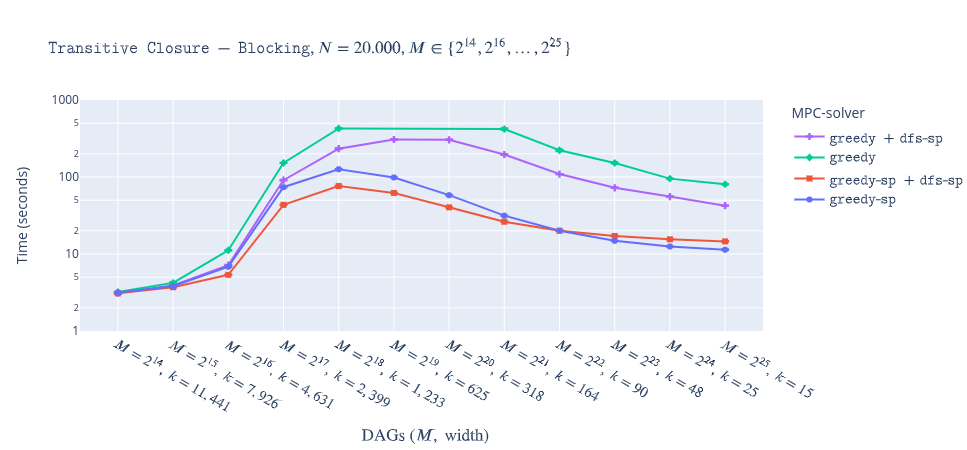}
    \caption{Running time of solver \texttt{Blocking greedy} in dataset \texttt{Transitive Closure} with different combinations of pre-processings \texttt{dfs-sp} and \texttt{greedy-sp}. Note the log-scale in the y-axis.}\label{fig:transitive_closure}
\end{figure*}

\subsection{Results}
\Cref{fig:random_dag} shows the running time of the MPC-solvers on the \texttt{Random DAG} dataset. Solvers starting from a \texttt{naive} solution are depicted with a solid line joining the corresponding data points\footnote{We consider \texttt{k3} and \texttt{k2} ``to start from a naive solution'' since, at each step, these consider the next vertex as a single path.}. The maximum flow-based solvers show a polynomial dependency in the number of edges of the input graph, with \texttt{Blocking naive} being the fastest (as predicted by theory as it uses a faster flow algorithm) followed by \texttt{DFS naive} and then by \texttt{BFS naive}. These results suggest that 1) the more complex \texttt{Blocking} algorithm pays off, as each step significantly reduces the path cover size, and that 2) although \texttt{BFS} ensures a polynomial running time for maximum flow, in the case of MPC this is unnecessary as $k \le |V|$ and \texttt{DFS} performs better in practice as residual paths are quickly found. Moreover, these solvers run out of time ($> 10$ mins.) after $|E| = 2^{24}, 2^{20}$ and $2^{18}$, respectively. Solvers starting from a \texttt{greedy} solution are shown with a dashed line. The solvers show a much faster running time, which stands below the $2$ mins. irrespective of the number of edges. As such, on dense graphs, these approaches are orders-of-magnitude faster that their \texttt{naive} counterparts. In this case, the difference between the different solvers is subtle, as substantially less residual paths must be found to transform the $O(\log{|V|})$-approximation to an MPC, and indeed \texttt{DFS} beats the machinery of \texttt{Blocking} from $|E| \ge 2^{22}$ and $k \le 494$. The parameterized linear time solvers \texttt{k3} and \texttt{k2} show a surprising running time behavior, which decreases with the number of edges. This behavior can be explained by the linear dependency on the number of edges: these algorithms process the edges, in constant-time, only during the initial edge sparsification, whereas vertices are charged with all the remaining machinery of the approach ($k^3$ and $k^2$ each, but also the associated constants). Both solvers run on less than $2$ mins. on every graph, and outperform the maximum flow-based solvers on dense graphs, from $|E| \ge 2^{20}$ in the case of \texttt{k2} and from $|E| \ge 2^{23}$ in the case of \texttt{k3}, being almost two orders-of-magnitude faster on the densest instance ($|E| = 2^{27}$). Finally, it is worth mentioning that \texttt{k2} outperforms \texttt{k3} on every instance tested, which shows that the more complex routines of the \texttt{k2} algorithm, as implemented in this work, manage to effectively shave a factor $k$ from the running time. In practical terms, the time saved by avoiding the full decomposition is larger that the time required to perform these savings.

\Cref{fig:path_partition} shows the running time of the solvers on the \texttt{Path Partition} dataset. The picture is very similar to the \texttt{Random DAG} dataset. In this case, \texttt{DFS} beats \texttt{Blocking} in sparse and very sparse graphs in \texttt{greedy} and \texttt{naive}, respectively. For all graphs in the dataset, the solvers \texttt{k3} and \texttt{k2} run in no more than $15$ secs. and $5$ secs., respectively.\\

\noindent\textbf{Pre-processing.} 
For \texttt{k3} and \texttt{k2}, using a pre-processing edge sparsification is counterproductive since these approaches perform their own sparsification as part of their routines. In the case of the maximum flow-based solvers the relative effect of the pre-processings is analogous, and thus we only comment on one solver.

\Cref{fig:transitive_closure} shows the running time of \texttt{Blocking greedy} on the \texttt{Transitive closure} dataset and different combinations of pre-processings \texttt{dfs-sp} and \texttt{greedy-sp}. All instances, except \texttt{greedy} at $M \in \{2^{19}, 2^{20}\}$, finish within the time limit of $10$ mins. We note that for $M \ge 2^{19}$ the number of edges in the corresponding graphs is larger than in the densest instance of the previous datasets, as such we call these graphs dense. On dense graphs, \texttt{dfs-sp} roughly decreases the running time in half, while \texttt{greedy-sp} reduces the running time by one order-of-magnitude. When using both heuristics \texttt{greedy-sp} and \texttt{dfs-sp}, we perceive a combined positive effect until $k \le 164$. For smaller values of $k$, performing both sparsifications does not pay off as \texttt{greedy-sp} is able to sparsify more edges (recall that \texttt{greedy-sp} sparsifies the edges to $|E'| = O(k|V|\log{|V|})$), but it does not affect the running time significantly either. On non-dense graphs ($M < 2^{19}$), applying both sparsifications dominates and it is up to $4$ times faster than plain \texttt{greedy}.

\section{Conclusions and Future Work}

We presented the first high-performance implementation of state-of-the-art MPC algorithms and showed that approaches parameterized by the width dominate the practical performance landscape on different kinds of random graphs. In particular, the parameterized linear time algorithms~\cite{caceres2022sparsifying,caceres2022minimum} shine on small-width instances, being orders-of-magnitude faster.  A theoretical open problem is whether there exists a linear time parameterized algorithm with a smaller dependency on $k$, in particular, is there and FPT-optimal algorithm running in  $O(k|V|+|E|)$ time? Recent works~\cite{kogan2023faster,caceres2023minimum} circumvent the $\Omega(k|V|)$ term\footnote{There are instances with $||\pathcover|| = \Omega(k|V|)$~\cite{caceres2023minimum}.} by computing a minimum chain cover whose size can be $o(k|V|)$. In practice, it is interesting to test whether these algorithms are effectively faster than our MPC-solvers or if these ideas can be used to improve the performance of our implementations. We also presented two new pre-processing heuristics based on transitive sparsification and showed how they improve the running time by an order-of-magnitude.

As mentioned in the introduction, an important application of MPC is reachability. In particular, it is known (see e.g.~\cite{kritikakis2023fast}) how to compute a constant-time reachability index of size $O(k|V|)$ in time $O(k|E'|)$, where $|E'|$ is the number of edges in the sparsest transitive sparsification, also known as transitive reduction. This result directly derives parameterized linear time solutions for the problems of constant-time reachability, transitive closure and transitive reduction, which can be implemented and compared against state-of-the-art solutions for those problems.

Finally, one algorithm we have no implemented is the $O(k^2|V|\log{|V|} + |E|)$-time approach of C\'aceres et~al.~\cite[Theorem 1.1]{caceres2022sparsifying} as this was later outperformed by the $O(k^2|V|+|E|)$-time algorithm~\cite{caceres2022minimum}. However, this divide-and-conquer approach is simple to parallelize~\cite[Theorem 1.2]{caceres2022sparsifying} and thus it could outperform our implementations when run on multiple processors.

\bibliographystyle{plain}
\bibliography{references}

\begin{thebibliography}{10}

\bibitem{ahujia1993network}
Ravindra~K Ahujia, Thomas~L Magnanti, and James~B Orlin.
\newblock {Network flows: Theory, algorithms and applications}.
\newblock {\em New Jersey: Prentice-Hall}, 1993.

\bibitem{barak1984maximal}
Amnon~B Barak and Paul Erd{\"o}s.
\newblock On the maximal number of strongly independent vertices in a random
  acyclic directed graph.
\newblock {\em SIAM Journal on Algebraic Discrete Methods}, 5(4):508--514,
  1984.

\bibitem{becker2016novel}
Ruben Becker, Maximilian Fickert, and Andreas Karrenbauer.
\newblock A novel dual ascent algorithm for solving the min-cost flow problem.
\newblock In {\em 2016 Proceedings of the Eighteenth Workshop on Algorithm
  Engineering and Experiments (ALENEX)}, pages 151--159. SIAM, 2016.

\bibitem{bova2015model}
Simone Bova, Robert Ganian, and Stefan Szeider.
\newblock Model checking existential logic on partially ordered sets.
\newblock {\em ACM Transactions on Computational Logic}, 17(2):1--35, 2015.

\bibitem{bunte2009overview}
Stefan Bunte and Natalia Kliewer.
\newblock An overview on vehicle scheduling models.
\newblock {\em Public Transport}, 1(4):299--317, 2009.

\bibitem{caceres2023minimum}
Manuel C{\'{a}}ceres.
\newblock Minimum chain cover in almost linear time.
\newblock In {\em Proceeding of the 50th International Colloquium on Automata,
  Languages, and Programming (ICALP 2023)}, volume 261, pages 31:1--31:12.
  Schloss Dagstuhl - Leibniz-Zentrum f{\"{u}}r Informatik, 2023.

\bibitem{caceres2022minimum}
Manuel Caceres, Massimo Cairo, Brendan Mumey, Romeo Rizzi, and Alexandru~I
  Tomescu.
\newblock Minimum path cover in parameterized linear time.
\newblock {\em arXiv preprint arXiv:2211.09659}, 2022.

\bibitem{caceres2022sparsifying}
Manuel C{\'a}ceres, Massimo Cairo, Brendan Mumey, Romeo Rizzi, and Alexandru~I
  Tomescu.
\newblock Sparsifying, shrinking and splicing for minimum path cover in
  parameterized linear time.
\newblock In {\em Proceedings of the 33rd Annual ACM-SIAM Symposium on Discrete
  Algorithms (SODA 2022)}, pages 359--376. SIAM, 2022.

\bibitem{CJ23}
Ghanshyam Chandra and Chirag Jain.
\newblock Sequence to graph alignment using gap-sensitive co-linear chaining.
\newblock In {\em Proceedings of the 27th Annual International Conference on
  Research in Computational Molecular Biology (RECOMB 2023)}, pages 58--73.
  Springer, 2023.

\bibitem{chang2015bridger}
Zheng Chang, Guojun Li, Juntao Liu, Yu~Zhang, Cody Ashby, Deli Liu, Carole~L
  Cramer, and Xiuzhen Huang.
\newblock Bridger: a new framework for de novo transcriptome assembly using
  {RNA}-seq data.
\newblock {\em Genome Biology}, 16(1):1--10, 2015.

\bibitem{chen2022maximum}
Li~Chen, Rasmus Kyng, Yang~P Liu, Richard Peng, Maximilian~Probst Gutenberg,
  and Sushant Sachdeva.
\newblock Maximum flow and minimum-cost flow in almost-linear time.
\newblock In {\em Proceedings of the 63rd IEEE Annual Symposium on Foundations
  of Computer Science (FOCS 2022)}, pages 612--623. IEEE, 2022.

\bibitem{chvatal1979greedy}
Vasek Chvatal.
\newblock A greedy heuristic for the set-covering problem.
\newblock {\em Mathematics of operations research}, 4(3):233--235, 1979.

\bibitem{ciurea2004sequential}
Eleonor Ciurea and Laura Ciupala.
\newblock Sequential and parallel algorithms for minimum flows.
\newblock {\em Journal of Applied Mathematics and Computing}, 15(1-2):53--75,
  2004.

\bibitem{colbourn1985minimizing}
Charles~J Colbourn and William~R Pulleyblank.
\newblock Minimizing setups in ordered sets of fixed width.
\newblock {\em Order}, 1(3):225--229, 1985.

\bibitem{cotumaccio2021indexing}
Nicola Cotumaccio and Nicola Prezza.
\newblock On indexing and compressing finite automata.
\newblock In {\em Proceedings of the 32nd ACM-SIAM Symposium on Discrete
  Algorithms (SODA 2021)}, pages 2585--2599. SIAM, 2021.

\bibitem{desrosiers1995time}
Jacques Desrosiers, Yvan Dumas, Marius~M Solomon, and Fran{\c{c}}ois Soumis.
\newblock Time constrained routing and scheduling.
\newblock {\em Handbooks in Operations Research and Management Science},
  8:35--139, 1995.

\bibitem{dezsHo2011lemon}
Bal{\'a}zs Dezs{\H{o}}, Alp{\'a}r J{\"u}ttner, and P{\'e}ter Kov{\'a}cs.
\newblock {LEMON}--an open source {C++} graph template library.
\newblock {\em Electronic Notes in Theoretical Computer Science},
  264(5):23--45, 2011.

\bibitem{dilworth1987decomposition}
Robert~P Dilworth.
\newblock A decomposition theorem for partially ordered sets.
\newblock {\em Classic Papers in Combinatorics}, pages 139--144, 1987.

\bibitem{dinitz2006dinitz}
Yefim Dinitz.
\newblock Dinitz’ algorithm: The original version and even’s version.
\newblock In {\em Theoretical Computer Science: Essays in Memory of Shimon
  Even}, pages 218--240. Springer, 2006.

\bibitem{edmonds1972theoretical}
Jack Edmonds and Richard~M Karp.
\newblock Theoretical improvements in algorithmic efficiency for network flow
  problems.
\newblock {\em Journal of the {ACM}}, 19(2):248--264, 1972.

\bibitem{eriksson2008viral}
Nicholas Eriksson, Lior Pachter, Yumi Mitsuya, Soo-Yon Rhee, Chunlin Wang,
  Baback Gharizadeh, Mostafa Ronaghi, Robert~W Shafer, and Niko Beerenwinkel.
\newblock Viral population estimation using pyrosequencing.
\newblock {\em {PLoS} Computational Biology}, 4(5):e1000074, 2008.

\bibitem{felsner2003recognition}
Stefan Felsner, Vijay Raghavan, and Jeremy Spinrad.
\newblock Recognition algorithms for orders of small width and graphs of small
  {D}ilworth number.
\newblock {\em Order}, 20(4):351--364, 2003.

\bibitem{ford1956maximal}
Lester~Randolph Ford and Delbert~R Fulkerson.
\newblock Maximal flow through a network.
\newblock {\em Canadian Journal of Mathematics}, 8:399--404, 1956.

\bibitem{fulkerson1956note}
Delbert~R Fulkerson.
\newblock Note on {D}ilworth's decomposition theorem for partially ordered
  sets.
\newblock {\em Proceedings of the American Mathematical Society},
  7(4):701--702, 1956.

\bibitem{gajarsky2015fo}
Jakub Gajarsk{\`y}, Petr Hlinen{\`y}, Daniel Lokshtanov, Jan Obdralek,
  Sebastian Ordyniak, MS~Ramanujan, and Saket Saurabh.
\newblock {FO} model checking on posets of bounded width.
\newblock In {\em Proceedings of the 56th IEEE 56th Annual Symposium on
  Foundations of Computer Science (FOCS 2015)}, pages 963--974. IEEE, 2015.

\bibitem{ikiz2006efficient}
Selma Ikiz and Vijay~K Garg.
\newblock Efficient incremental optimal chain partition of distributed program
  traces.
\newblock In {\em Proceedings of the 26th IEEE International Conference on
  Distributed Computing Systems (ICDCS 2006)}, pages 18--18. IEEE, 2006.

\bibitem{Jagadish90}
H.~V. Jagadish.
\newblock A compression technique to materialize transitive closure.
\newblock {\em ACM Transactions on Database Systems}, 15(4):558--598, 1990.

\bibitem{jaskowski2011formal}
Wojciech Ja{\'s}kowski and Krzysztof Krawiec.
\newblock Formal analysis, hardness, and algorithms for extracting internal
  structure of test-based problems.
\newblock {\em Evolutionary Computation}, 19(4):639--671, 2011.

\bibitem{kahn1962topological}
Arthur~B Kahn.
\newblock Topological sorting of large networks.
\newblock {\em Communications of the {ACM}}, 5(11):558--562, 1962.

\bibitem{kogan2022beating}
Shimon Kogan and Merav Parter.
\newblock Beating matrix multiplication for $n^{1/3}$-directed shortcuts.
\newblock In {\em 49th International Colloquium on Automata, Languages, and
  Programming (ICALP 2022)}. Schloss Dagstuhl-Leibniz-Zentrum f{\"u}r
  Informatik, 2022.
\newblock Full version available at
  \url{https:\\www.weizmann.ac.il/math/parter/sites/math.parter/files/uploads/main-lipics-full-version_3.pdf}.

\bibitem{kogan2023faster}
Shimon Kogan and Merav Parter.
\newblock Faster and unified algorithms for diameter reducing shortcuts and
  minimum chain covers.
\newblock In {\em Proceedings of the 34th Annual ACM-SIAM Symposium on Discrete
  Algorithms (SODA 2023)}, pages 212--239. SIAM, 2023.

\bibitem{kovacs2015minimum}
P{\'e}ter Kov{\'a}cs.
\newblock Minimum-cost flow algorithms: an experimental evaluation.
\newblock {\em Optimization Methods and Software}, 30(1):94--127, 2015.

\bibitem{kowaluk2008path}
Miros{\l}aw Kowaluk, Andrzej Lingas, and Johannes Nowak.
\newblock A path cover technique for lcas in dags.
\newblock In {\em Scandinavian Workshop on Algorithm Theory}, pages 222--233.
  Springer, 2008.

\bibitem{kritikakis2023fast}
Giorgos Kritikakis and Ioannis~G Tollis.
\newblock Fast reachability using {DAG} decomposition.
\newblock In {\em Proceedings of the 21st International Symposium on
  Experimental Algorithms (SEA 2023)}. Schloss Dagstuhl-Leibniz-Zentrum f{\"u}r
  Informatik, 2023.

\bibitem{ma2023chaining}
Jun Ma, Manuel Cáceres, Leena Salmela, Veli Mäkinen, and Alexandru~I Tomescu.
\newblock {Chaining for Accurate Alignment of Erroneous Long Reads to Acyclic
  Variation Graphs}.
\newblock {\em Bioinformatics}, page btad460, 07 2023.

\bibitem{mackinnon1985optimal}
Stephen~J MacKinnon, Peter~D Taylor, Henk Meijer, and Selim~G. Akl.
\newblock An optimal algorithm for assigning cryptographic keys to control
  access in a hierarchy.
\newblock {\em IEEE Transactions on Computers}, 34(09):797--802, 1985.

\bibitem{makinen2019sparse}
Veli M{\"a}kinen, Alexandru~I Tomescu, Anna Kuosmanen, Topi Paavilainen, Travis
  Gagie, and Rayan Chikhi.
\newblock Sparse dynamic programming on {DAG}s with small width.
\newblock {\em ACM Transactions on Algorithms}, 15(2):1--21, 2019.

\bibitem{ntafos1979path}
Simeon~C Ntafos and S~Louis Hakimi.
\newblock On path cover problems in digraphs and applications to program
  testing.
\newblock {\em IEEE Transactions on Software Engineering}, 5(5):520--529, 1979.

\bibitem{simon1988improved}
Klaus Simon.
\newblock An improved algorithm for transitive closure on acyclic digraphs.
\newblock {\em Theoretical Computer Science}, 58(1-3):325--346, 1988.

\bibitem{tarjan1976edge}
Robert~E Tarjan.
\newblock Edge-disjoint spanning trees and depth-first search.
\newblock {\em Acta Informatica}, 6(2):171--185, 1976.

\bibitem{tomlinson1997monitoring}
Alexander~I Tomlinson and Vijay~K Garg.
\newblock Monitoring functions on global states of distributed programs.
\newblock {\em Journal of Parallel and Distributed Computing}, 41(2):173--189,
  1997.

\bibitem{trapnell2010transcript}
Cole Trapnell, Brian~A Williams, Geo Pertea, Ali Mortazavi, Gordon Kwan,
  Marijke~J Van~Baren, Steven~L Salzberg, Barbara~J Wold, and Lior Pachter.
\newblock Transcript assembly and quantification by {RNA-S}eq reveals
  unannotated transcripts and isoform switching during cell differentiation.
\newblock {\em Nature Biotechnology}, 28(5):511, 2010.

\bibitem{zhan2016graph}
Xianyuan Zhan, Xinwu Qian, and Satish~V Ukkusuri.
\newblock A graph-based approach to measuring the efficiency of an urban taxi
  service system.
\newblock {\em IEEE Transactions on Intelligent Transportation Systems},
  17(9):2479--2489, 2016.

\end{thebibliography}
\end{document}